\documentclass[12pt]{article}
\title{Limit Cycle Oscillations, response time and the time-dependent solution to the Lotka-Volterra Predator-Prey model}
\author{M. Leconte$^{1,a}$, P. Masson$^2$ and Lei Qi$^1$ \\
$^1$ Korea Institute of Fusion Energy (KFE), Daejeon 34133, South Korea \\
$^2$ Independent researcher, https://analyticphysics.com \\
 \quad \\
$^a$ Author to whom correspondence should be addressed: mleconte@kfe.re.ke 
}

\usepackage{graphicx}
\usepackage{amsmath}
\usepackage{hyperref}

\begin{document}
\maketitle

\newcommand{\g}{\gamma}

\newcommand{\lambdaoneprey}{\Lambda_{-1}^{\rm prey}}
\newcommand{\lambdazprey}{\Lambda_0^{\rm prey}}
\newcommand{\lambdaonepreda}{\Lambda_{-1}^{\rm preda}}
\newcommand{\lambdazpreda}{\Lambda_0^{\rm preda}}

\newcommand{\lambdajprey}{\Lambda_j^{\rm prey}}
\newcommand{\lambdajpreda}{\Lambda_j^{\rm preda}}

\begin{abstract}
In this work, the time-dependent solution for the Lotka-Volterra Predator-Prey model is derived with the help of the Lambert W function. This allows an exact analytical expression for the period of the associated limit-cycle oscillations (LCO),  and also for the response time between predator and prey population. These results are applied to the predator-prey interaction of zonal density corrugations and turbulent particle flux in gyrokinetic simulations of collisionless trapped-electron model (CTEM) turbulence. In the turbulence simulations, the response time is shown to increase when approaching the linear threshold, and the same trend is observed in the Lotka-Volterra model.
\end{abstract}

\section{Introduction}
Limit-cycle oscillations occur in many areas of Biology and Physics \cite{Murray1989,CrossGreenside2009}. One popular model to describe these phenomena is the Lotka-Volterra model \cite{Murray1989, scholarpedia-LV} and its extensions. Lotka \cite{Lotka1920} and Volterra \cite{Volterra1926} derived it independently to describe the nonlinear interaction of predator and prey populations. Therefore, it is now widely known as the `Predator-Prey model'. In Physics, many nonlinear interactions can be described with this model. In Plasma Physics, on which we focus here, it describes the interaction between axisymmetric self-generated flows (zonal flows), which acts as the predator, and the microturbulence - the prey - that drives them \cite{Diamond1994}. In the plasma turbulence context, this model is a building-block for more extended models of the Low to High confinement (L-H) transition in fusion plasmas such as the Kim-Diamond model \cite{KimDiamond2003}. See e.g. Ref. \cite{DiamondHasegawaMima2011} for a review. Refs. \cite{ZhuChapmanDendy2013} and \cite{DamBronsRasmussen2013} analyzed such extended predator-prey models - which are not integrable - using bifurcation theory. The limit-cycles associated to the Lotka-Volterra model have one particular feature that is interesting: They exhibit multiple time-scales. Hence, such models may be useful to understand certain types of relaxation-oscillations, and intermittent transport in turbulent plasmas. The well-known Van-der-Pol oscillator \cite{VanderPol1926} and the closely-related Rayleigh oscillator \cite{Rayleigh1883} are also examples of a system showing relaxation-oscillations. During the L-H transition, a dithering phase (also called Intermittent phase or I-phase) is often observed experimentally when the heating power is slowly ramped up \cite{Schmitz2012, MikiDiamondGurcan2012}. It can be understood as limit-cycle oscillations (LCO) between turbulence energy and zonal flow energy, as observed in gyrokinetic simulations \cite{ZLin1999, SKobayashiGurcanDiamond2015}.
It is this feature that we focus on in this work.
More precisely, we focus on the response-time between these two quantities. Refs. \cite{Kosuga2013, Gurcan2013, Kosuga2014} showed - based on a different model - that this response-time is a key quantity to understand nonlinear interactions. In the simplified framework of Drift-wave turbulence, there are several candidates to explain the interaction between zonal density corrugations and the turbulence. One of the authors (M. Leconte) proposed a model based on the nonlinear modulation of the transport crossphase between density and potential perturbations \cite{LeconteKobayashi2021, LeconteSingh2019}. Another model, based on stochastic noise due to turbulence was proposed in Ref. \cite{SinghDiamond2021}. The main results of this article are: \\
(i) a new analytical solution opens the possibility of directly fitting experimental data of LCO to extract its key predator-prey features, with only two fitting parameters, c.f. Eqs. (\ref{def-LV-integral1},\ref{anasol-predator},\ref{anasol-prey},\ref{def-LV-integral2}). \\
(ii) the response time between turbulence energy and zonal energy increases as marginality is approached, in collision-less trapped electron mode turbulence simulations. A similar trend is observed in the Lotka-Volterra model. \\
The rest of this article is organized as follows: In Section 2, we describe the Lotka-Volterra Predator-Prey model, and we derive its time-dependent solution analytically. In Section 3, we apply the newly-found solutions to understand the predator-prey dynamics between zonal density perturbations and the turbulent particle flux observed in global gyrokinetic simulations of collision-less trapped-electron mode turbulence (CTEM) \cite{LeiQi2019, LeiQi2021, LeiQi2017, LeiQi-inprep-2021}. Finally in Section 4, we discuss the results and give a conclusion.

\section{Model}
We consider the following Lotka-Volterra Predator-Prey model:
\begin{eqnarray}
\dot x & = & \g x - \alpha_1 x y
\label{ppx00} \\
\dot y & = & \alpha_2 xy - \mu y
\label{ppy00}
\end{eqnarray}
Here, $x$ denotes the prey population, and $y$ denotes the predator population, and $\dot {u} = du/dt$, with $u=x,y$, denotes the time derivative. The parameter $\g$ denotes the prey growth-rate (birth-rate) in the absence of predator, and $\mu$ is the predator damping rate (death-rate) in the absence of prey. The coefficients $\alpha_1, \alpha_2$ are positive constants.
In applications to plasma turbulence, the predator is usually taken as zonal flow energy, and the prey as turbulence energy \cite{Diamond1994, KimDiamond2003,DiamondHasegawaMima2011}. Here, we take the predator as zonal density corrugations driven by nonlinear modulation of the transport crossphase \cite{LeconteKobayashi2021} and the prey as turbulence energy. In section 3, this model will be applied to gyrokinetic simulations of CTEM turbulence. There are several candidate mechanisms for the nonlinear generation of zonal density by the turbulence \cite{LeconteKobayashi2021, SinghDiamond2021} based on fluid models, although a specific application to CTEM has not been proposed yet. For this reason, we treat the Lotka-Volterra as a phenomenological model here. However, deriving a predator-prey like reduced-model directly from the bounce-averaged gyrokinetic equation would be an important task. One could use a similar method as in Ref. \cite{Kosuga2014}, where a derivation of the traffic-jam model from the nonlinear gyrokinetic equation is sketched.

\subsection{Normalization of the model}
It is convenient to re-define the variables, so as to decrease the number of independent parameters \cite{PMasson2021}.
We make the following change of variables:
\begin{eqnarray}
X = \frac{\alpha_2}{\mu} x, \\
Y = \frac{\alpha_1}{\mu} y
\end{eqnarray}
Using this change of variables, one obtains after some algebra:
\begin{eqnarray}
\dot X & = & \frac{1}{\delta} X - XY,
\label{ppx1} \\
\dot Y & = & XY - Y,
\label{ppy1}
\end{eqnarray}
with $\delta = \mu / \g$, and where the time has been re-scaled to $\mu t \to t$. Note that for typical values of parameters, $\delta \ll 1$, but our analysis is valid for any value of $\delta$.

\subsection{Energy conservation}
\begin{figure}
\begin{center}
\includegraphics[width=0.5\linewidth]{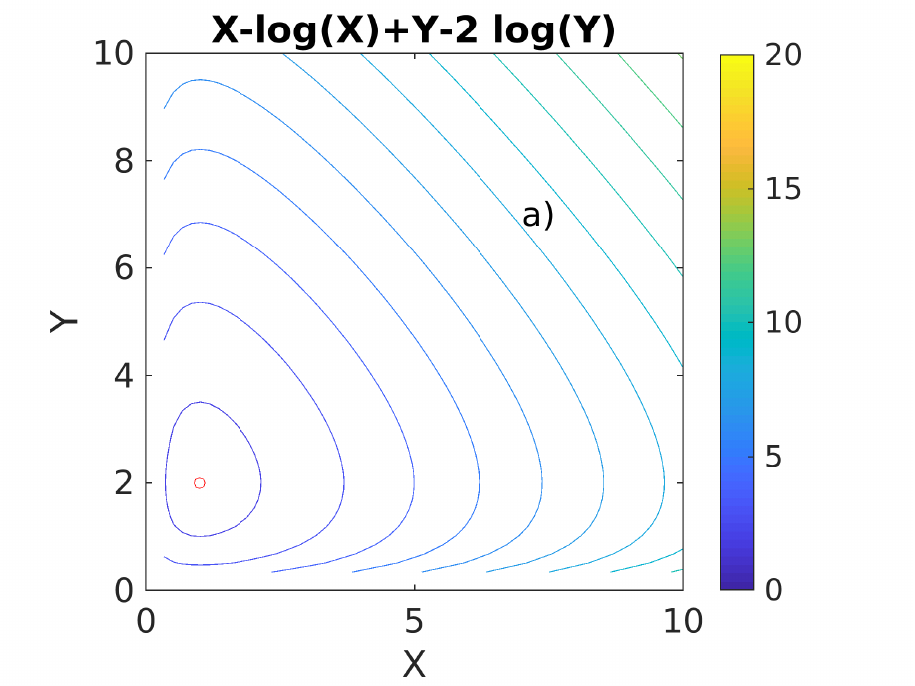}\includegraphics[width=0.5\linewidth]{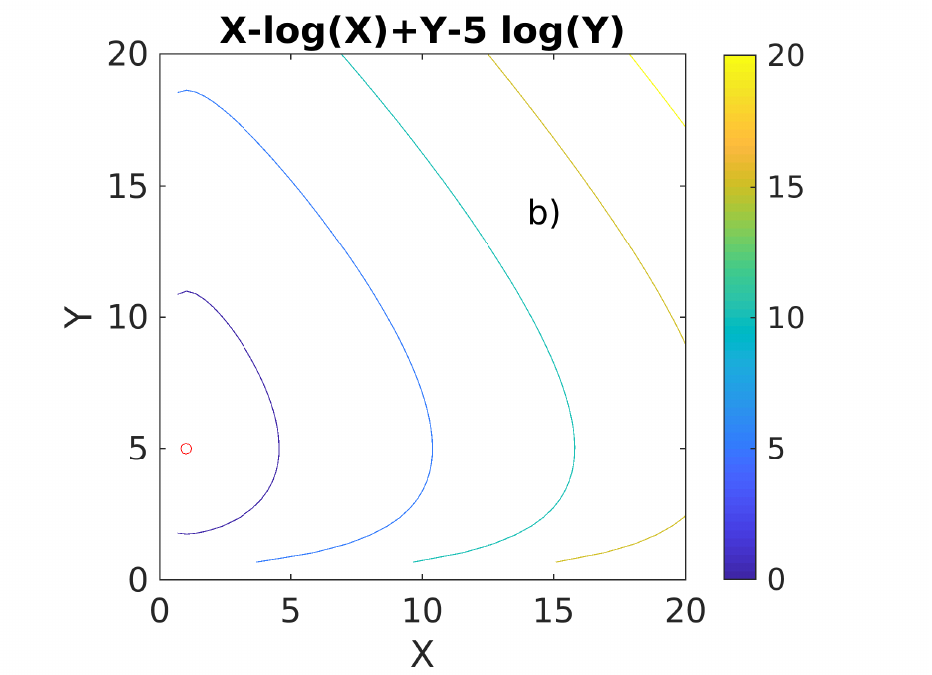}
\caption{Contours of the energy integral (\ref{hamilton1}) for the Lotka-Volterra system, for a) $\delta=0.5$, and b) $\delta=0.2$. The red circle indicates the center of the associated limit-cycles.}
\label{fig-hamilton}
\end{center}
\end{figure}

\begin{figure}
\begin{center}
\includegraphics[width=0.5\linewidth]{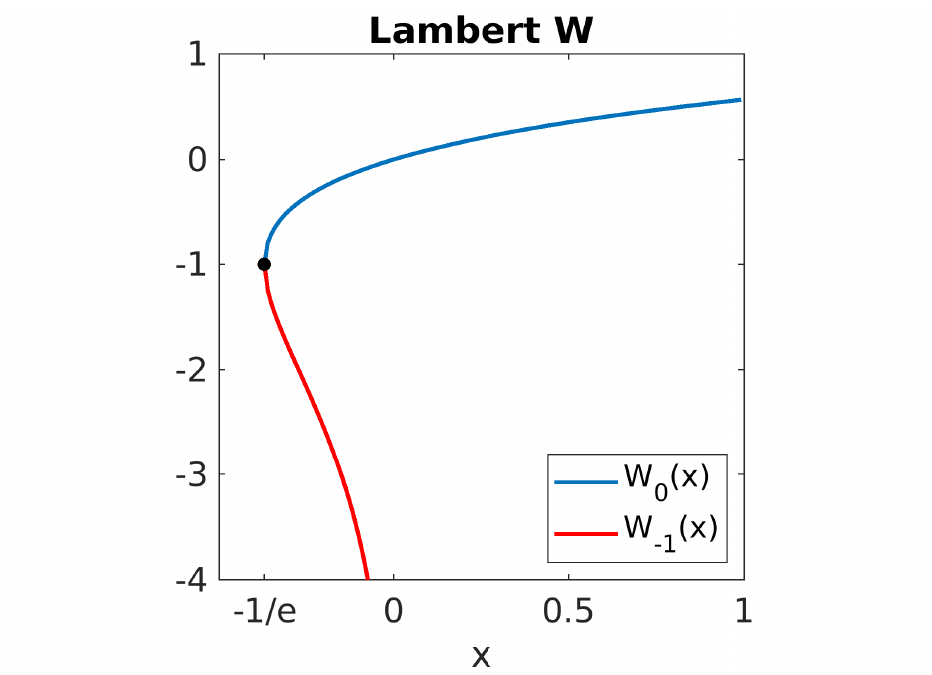}
\end{center}
\caption{Plot of the Lambert W function: $W_j(x), j=0,-1$.}
\label{fig-Lambert}
\end{figure}

It is well-known that the system (\ref{ppx1},\ref{ppy1}) has an invariant associated to its limit-cycle. Here, we briefly review the derivation of this invariant (a Lyapunov function). Dividing Eq. (\ref{ppy1}) by Eq. (\ref{ppx1}), one obtains:
\begin{equation}
\frac{dY}{dX} = - \frac{Y}{X} \cdot \frac{1 - X}{\frac{1}{\delta}- Y}
\end{equation}
Since this is a separable ordinary differential equation, one obtains - after some algebra - the following energy integral:
\begin{equation}
X -  \ln X + Y - \frac{1}{\delta} \ln Y = E,
\label{hamilton1}
\end{equation}
where $E= {\rm Cst}$ is the total energy determined by initial conditions $X(t=0)$ and $Y(t=0)$. It can be shown that this quantity is actually a generalized Hamiltonian \cite{Duarte1998}. Contours of the Hamiltonian are shown for two values of the parameter $\delta=0.5$ and $\delta=0.2$ [Fig. \ref{fig-hamilton}]. For small values of $\delta$ [Fig. \ref{fig-hamilton}b], one observes that the limit cycles become more elongated in the $Y$ direction, i.e. the predator population has a very large amplitude compared to the prey population.
Up to now, the nonlinear solutions to system (\ref{ppx1},\ref{ppy1}) are thus obtained in an implicit form, through their representation as a projection of the 4D dynamical phase-space $(X, \dot X, Y, \dot Y)$ onto the 2D space $(X,Y)$. One can make the analogy with the Jacobi elliptic functions (c.f. Appendix). In the following, we will go one step further, to obtain the nonlinear time-dependent solutions in explicit form. \\
It was shown in Ref. \cite{PMasson2021} that the energy integral can be used to express either of the variable $X$ or $Y$ in terms of the other, using the Lambert W function \cite{Corless1996}. Note that this result was also obtained independently in the latter Reference (page 336 of \cite{Corless1996}). This function is solution to the transcendental equation: $W e^{W}= u$. Applied to the energy integral Eq. (\ref{hamilton1}), one obtains after some algebra:
\begin{eqnarray}
X(Y) & = & -W_j( -Y^{- \frac{1}{\delta}} e^{Y-E} ), \\
Y(X) & = & - \frac{1}{\delta} W_j ( -\delta X^{- \delta} e^{\delta (X-E)} ),
\end{eqnarray}
where the subscript $j=0, -1$ denotes the relevant branch of the Lambert W function. The $W_0$ function is known as the \emph{principal branch}, while the $W_{-1}$ function is called \emph{negative branch}. They correspond to the two roots of the transcendental equation $W e^W = u$, for $u$ real-valued.
For clarity, the two branches of the Lambert W function are plotted [Fig. \ref{fig-Lambert}]

\section{Time-dependent solutions}
Let us first consider the solution for the predator population $Y(t)$.
Starting from the predator evolution  Eq. (\ref{ppy1}). we write it in the form:
\begin{equation}
dt = g(X,Y) dY
\label{predator1}
\end{equation}
Here, the function is given by $g(X,Y) = - \frac{1}{(1 - X)Y}$.
Now, we use Eq. (\ref{hamilton1}) to express the prey $X$ in terms of the predator $Y$ and total energy $E$:
\begin{equation}
X = - {\rm W}_j (- Y^{- \frac{1}{\delta}} e^{Y-E}),
\end{equation}
where $W_j$ denotes the Lambert W function \cite{Corless1996}, and $j=0, -1$ is the associated branch.

\begin{figure}
\begin{center}
\includegraphics[width=0.5\linewidth]{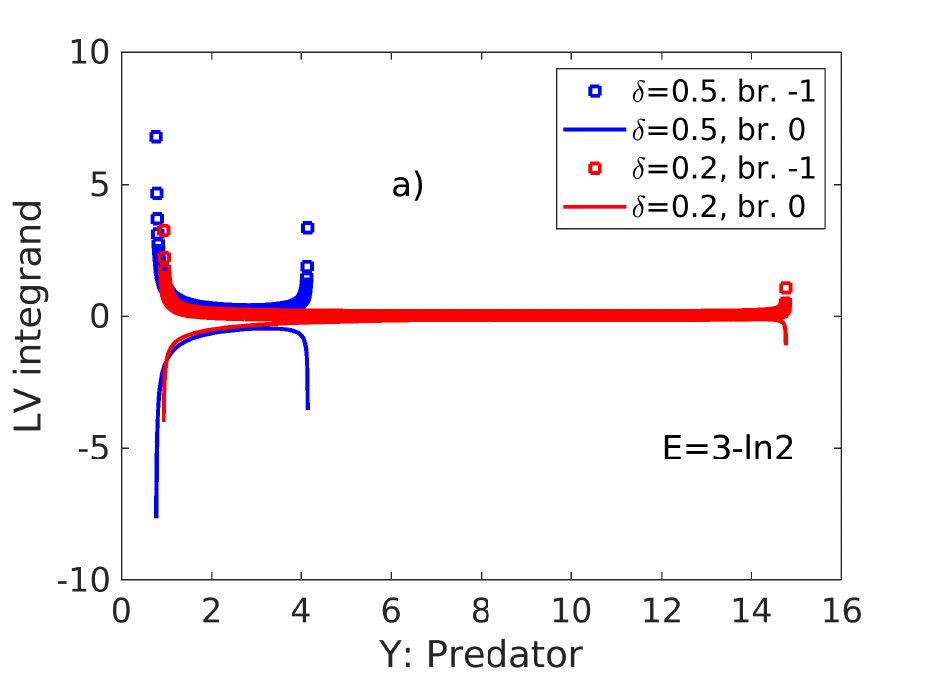}\includegraphics[width=0.5\linewidth]{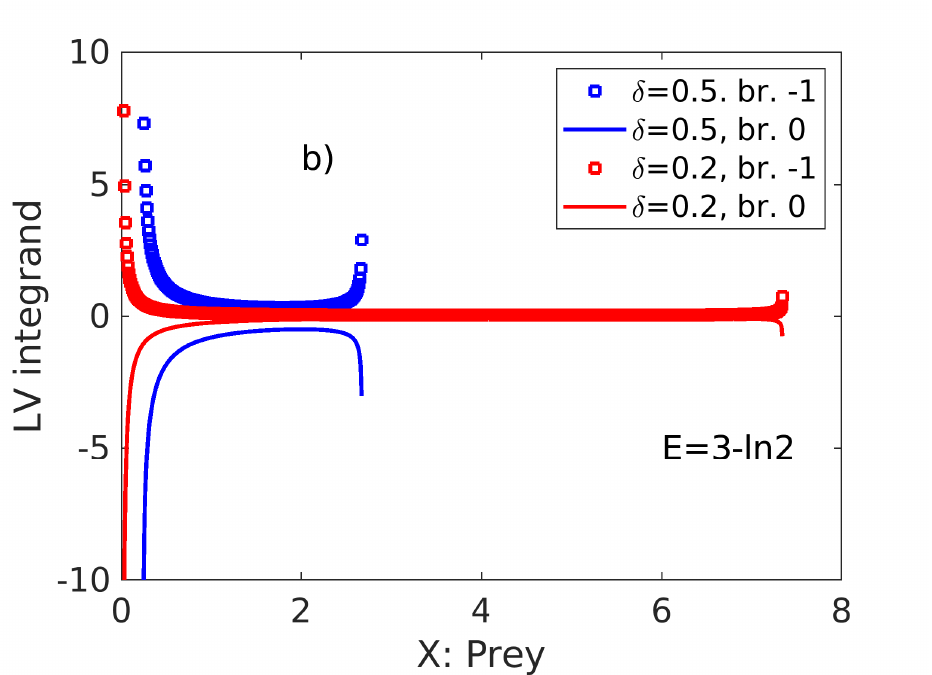}
\caption{Integrand of the a) second and b) first Lotka-Volterra integrals Eqs. (\ref{def-LV-integral1}) and  (\ref{def-LV-integral2}).}
\label{fig-LV-integrand}
\end{center}
\end{figure}

Integrating both sides of Eq. (\ref{predator1}) yields:
\begin{equation}
t = G_{LV}^j(Y,E)
\label{predator2}
\end{equation}
where we call $G_{LV}^j$ the `second Lotka-Volterra' integral, defined as:
\begin{equation}
G_{LV}^j(Y,E) = T_{Y_{\rm min}} - \int_{Y_{\rm min}}^{Y} \frac{dY'}{Y' [1 + {\rm W}_j (- Y'^{- \frac{1}{\delta}} e^{Y'-E})]},
\label{def-LV-integral1}
\end{equation}
where $W_j = W_0$ or $W_j=W_{-1}$, depending on the branch of the Lambert $W$ function considered.
The integrand of the Lotka-Volterra integral (\ref{def-LV-integral1}) for the branches j=0,-1 is shown v.s. $Y$ for different values of the parameter $\delta$, for an energy of $E=3-\ln2$ [Fig. \ref{fig-LV-integrand}]. This value of energy corresponds to the initial conditions $(X_0,Y_0)=(2,1)$. Note the vertical asymptotes corresponding to the minima and maxima of the predator population $Y$. The integration constant $T_{Y_{\rm min}}$ is given by expression (\ref{ap-tymin}) in Appendix.

Next, we invert Eq. (\ref{predator2}) to obtain the time-dependent solution
$Y(t,E) = \lambdajpreda(t,E)	$
for the predator population, where the function
$\lambdajpreda(t,E)$
is given by:
\begin{equation}
\lambdajpreda(t,E)	 = G_{LV}^{-1}(t,E),
\label{anasol-predator}
\end{equation}
where the superscript $^{-1}$ denotes the function inverse (not to be confused with the index $j=0,-1$ of the branch), and we use the shortcut notation $G_{LV}=G_{LV}^j$.
Care must be taken when inverting Eq. (\ref{predator2}), because the real-valued Lambert W function has two branches: $W_0(x)$ and $W_{-1}(x)$.
As a second step, one applies a similar procedure for the prey population X, $dt = f(X,Y) dX$, with $f(X,Y)= \frac{1}{(\frac{1}{\delta} -Y)X}$, to obtain the time-dependent solution
$X(t,E) = \lambdajprey(t,E)$
for the prey population:
\begin{equation}
\lambdajprey(t,E) =  F^{-1}_{LV}(t,E),
\label{anasol-prey}
\end{equation}
where the superscript $^{-1}$ denotes the function inverse, and we define the first Lotka-Volterra integral $F_{LV}$ as:
\begin{equation}
F_{LV}^j(X,E) = T_{X_{\rm min}} - \int_{X_{\rm min}}^{X} \frac{dX'}{\frac{1}{\delta} X' [1+ {\rm W}_j (\delta X'^{- \delta} e^{\delta(X'-E)})]},
\label{def-LV-integral2}
\end{equation}

\begin{figure}
\begin{center}
\includegraphics[width=0.5\linewidth]{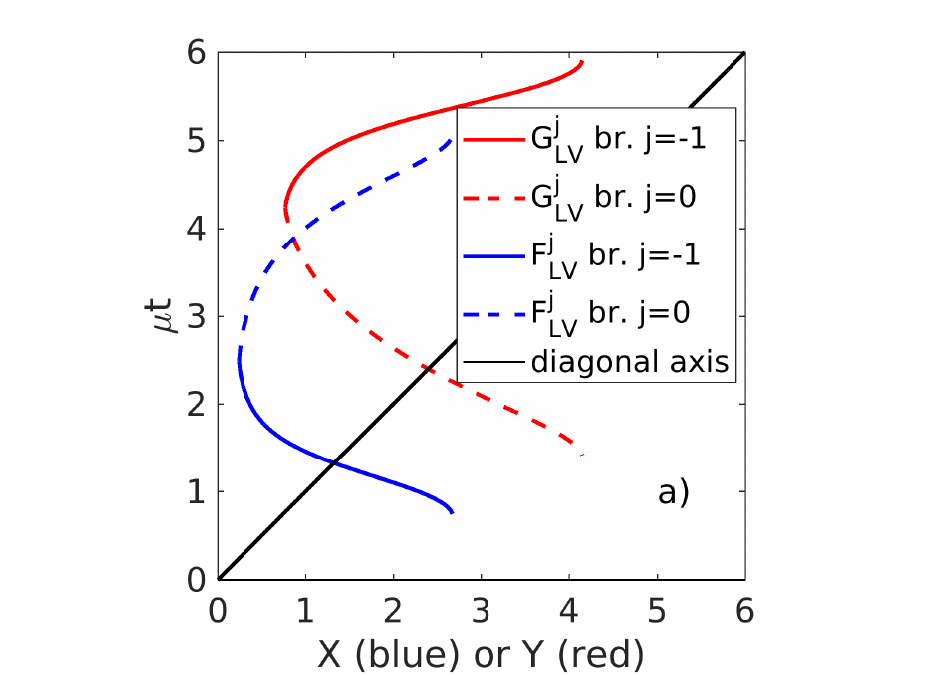}\includegraphics[width=0.5\linewidth]{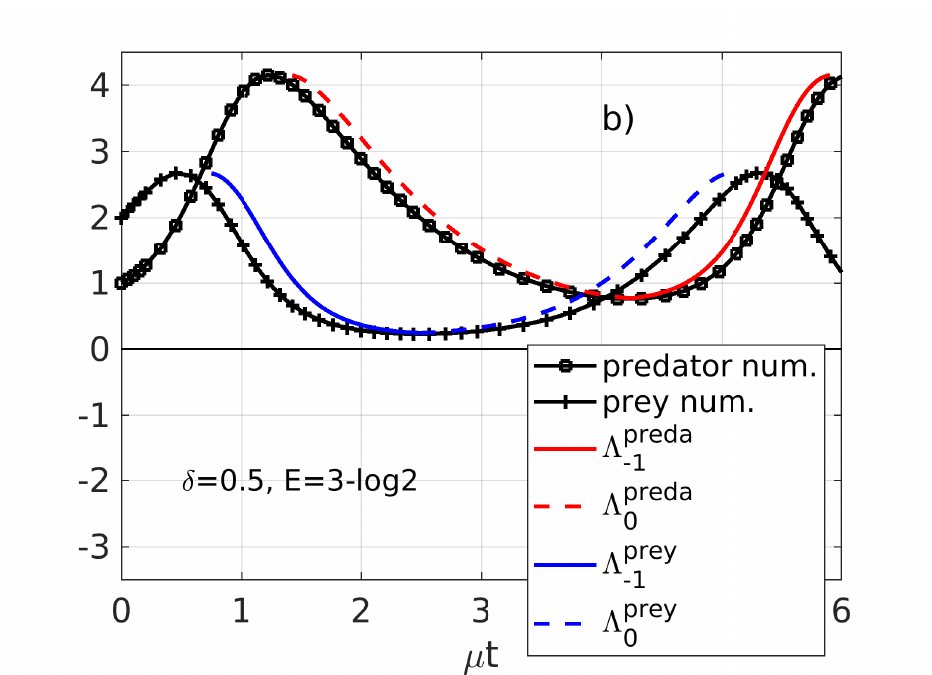}
\caption{a) Inverse functions $G_{LV}^j(Y,E)$ (red) and $F_{LV}^j(X,E)$ (blue) given by Eqs. (\ref{def-LV-integral1}) and (\ref{def-LV-integral2}), and b) Analytical solutions $Y(t)$ (red) and $X(t)$ (blue) of the Lotka-Volterra system, for an energy of $E=3-\ln 2 = 2.3069$, compared with numerical solutions (open symbols). The parameter is $\delta = 0.5$.}
\label{fig-LV-t-ana}
\end{center}
\end{figure}

We call the analytical solutions $\lambdajprey(t,E)$, and $\lambdajpreda(t,E)$ given by expressions (\ref{anasol-predator}) and (\ref{anasol-prey}) the `Lotka-Volterra functions'.
The result is shown for a value $\delta=0.5$ of the parameter and an energy of $E \simeq 2$ [Fig. \ref{fig-LV-t-ana}].

\section{Response-time comparison with gyrokinetic simulations of CTEM turbulence}

\begin{figure}
\begin{center}
\includegraphics[width=0.5\linewidth]{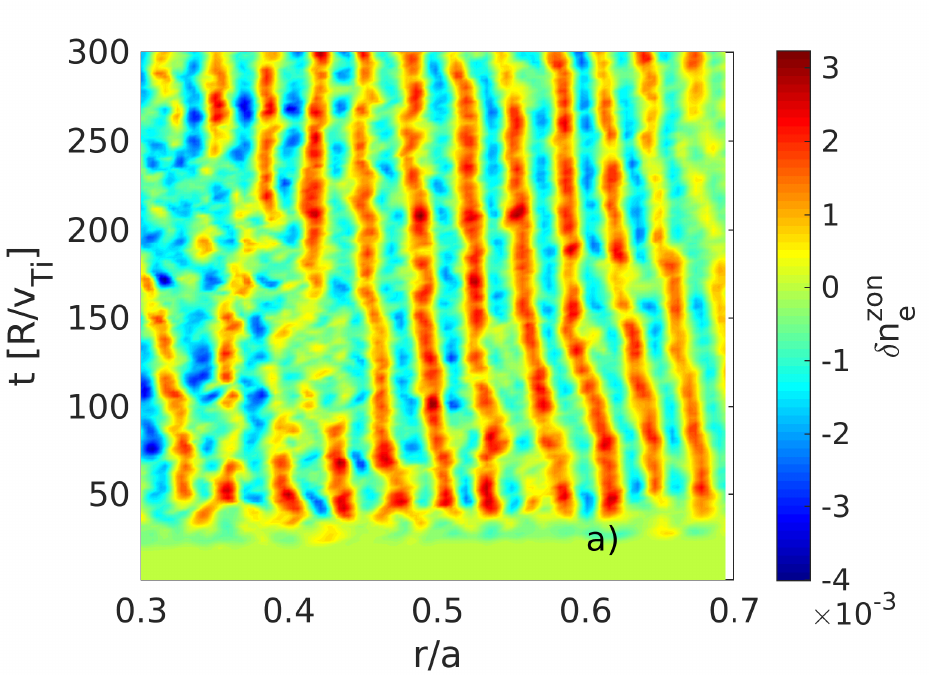}\includegraphics[width=0.5\linewidth]{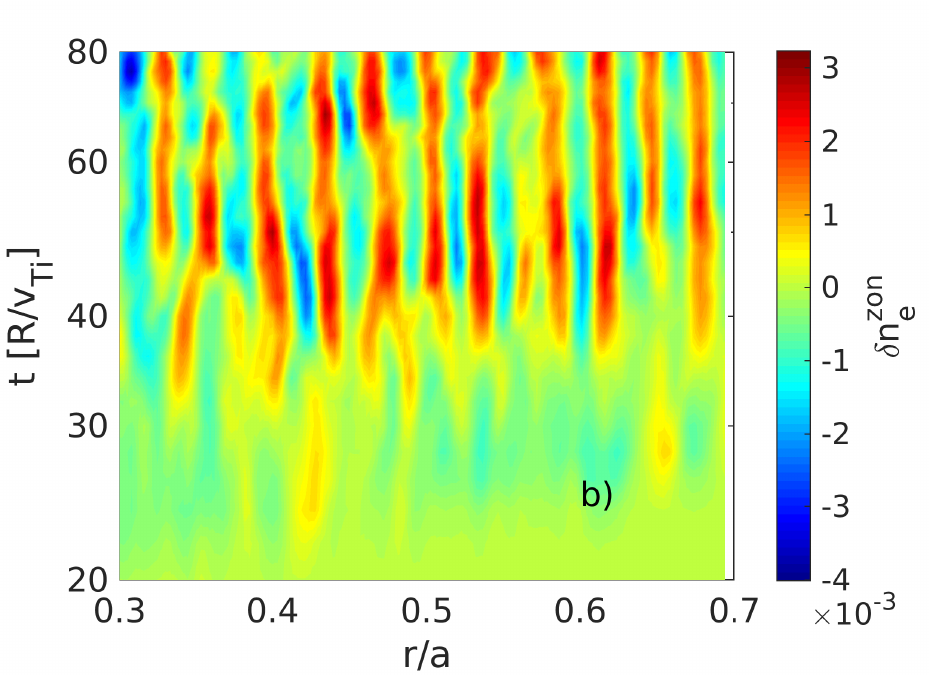}
\caption{a) Colormap of zonal electron density $\delta n_{ez}$ v.s. radius and time showing the zonal staircase pattern, and b) close-up in log-scale around the time of staircase formation $t=20-80~R/V_{Ti}$.}
\label{fig-ctem-staircase}
\end{center}
\end{figure}

\begin{figure}
\includegraphics[width=0.5\linewidth]{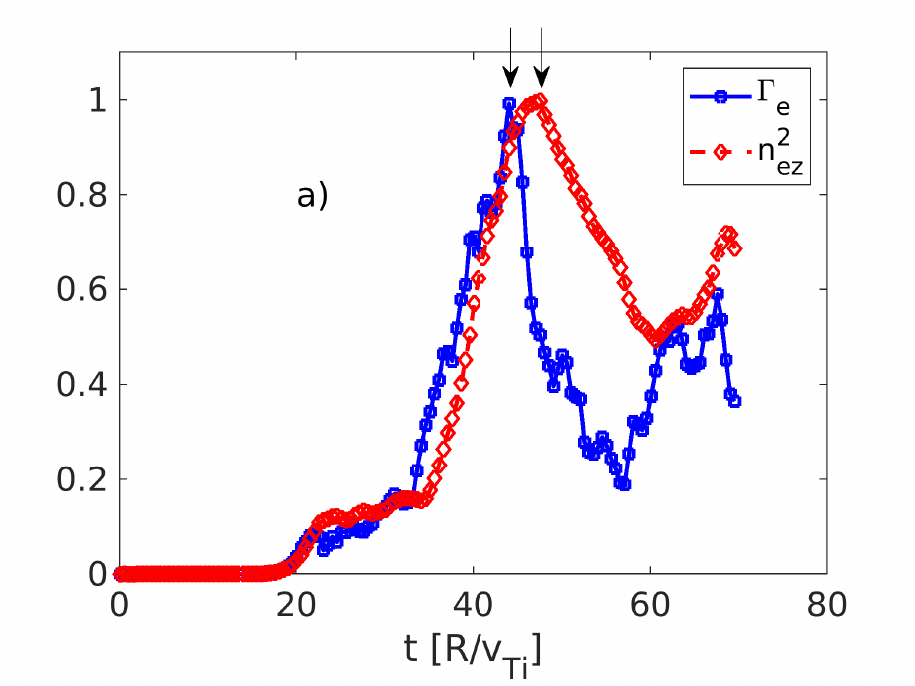}\includegraphics[width=0.5\linewidth]{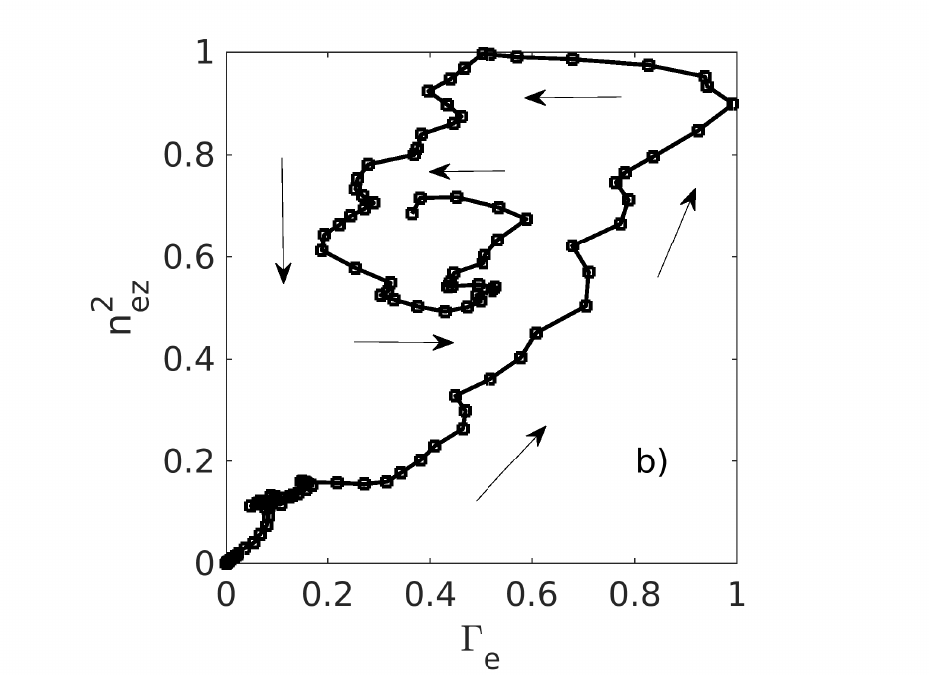}
\caption{a) Time-trace of electron particle flux $\Gamma_e(t)$ and zonal density energy $n_{ez}^2(t)$, and b) associated limit-cycle in dynamical phase-space ($\Gamma_e$,$n_{ez}^2$).}
\label{fig-ctem-sim}
\end{figure}

The collisionless trapped-electron mode (CTEM) is an instability due to the electron toroidal precession-drift resonance - a process similar to inverse Landau damping - in the low-collisionality regime \cite{Adam1976}. Its sources of energy are the electron temperature gradient and the density gradient. CTEM instability is in the ion-scale range, with a typical poloidal wave-number $k_\theta \rho_i \sim 1$, where $\rho_i = \sqrt{m_i T_i} / eB$ is the ion-gyroradius, $T_i$ is the ion temperature and other notations are standard.
From collisionless trapped-electron mode (CTEM) gyrokinetic simulations, zonal density perturbations $\delta n_{ez}$ and electron particle flux $\Gamma_e$ are obtained. Figure \ref{fig-ctem-staircase} shows the zonal density corrugations v.s. radius and time. One clearly observes the radially oscillating zonal pattern known as `staircase' \cite{Kosuga2013, Gurcan2013, Kosuga2014, LeiQi2019, LeiQi2021, MJChoi2019, DifPradalier2015, HornungDifPradalier2017, DifPradalier2017, AshourvanDiamond2017, GuoDiamondHughes2019,  WLiu2021, GarbetPanicoVarennes2021}. The time-trace of electron particle flux $\Gamma_e$ and zonal density energy $n_{ez}^2$ are shown [Fig \ref{fig-ctem-sim}a]. These quantities are spectral averages, e.g. $\Gamma_e(t) = \sum_{k_r\rho_i\in[0.4,1]} |\Gamma_e(t,k_r)|$, around the radial wavenumber $k_r \rho_i=0.78$, which is a characteristic scale of the zonal staircase pattern. The simulations are performed with the gyrokinetic code gKPSP \cite{Kwon2017} which solves the nonlinear gyrokinetic equations for ions \cite{Hahm1988} and bounce-averaged kinetics for trapped electrons \cite{FongHahm1999}.
In the simulations, the equilibrium gradients are $R/L_n =2.2$, $R/L_{Ti}=2.2$, and $R/L_{Te}=4.0-12.0$ where $n,T_e$ and $T_i$ denote the electron density, electron temperature and ion temperature, and e.g. $L_n = - [\frac{1}{n} \frac{dn}{dr}]^{-1}$. At mid-radius, the inverse aspect ratio is $r/R=0.18$, safety factor $q=1.4$, magnetic shear $\hat s = q'r/q=0.78$, $T_i=T_e$. Hydrogen is the main ion $m_i/m_e=1836$ and plasma elongation $\kappa=2$. The turbulent transport of this plasma is dominated by CTEM \cite{LeiQi2021, LeiQi2017}.
A limit-cycle type of dynamics between $\Gamma_e$ and $n_{ez}^2$ is clearly observed in dynamical phase-space [Fig \ref{fig-ctem-sim}b], although its amplitude decreases with time, probably due to additional turbulent dissipation. This is probably the reason for the spiraling in Fig \ref{fig-ctem-sim}b.The Lotka-Volterra model does not take into account this additional dissipation, but the overall dynamics is similar to the model.

\begin{figure}
\includegraphics[width=0.5\linewidth]{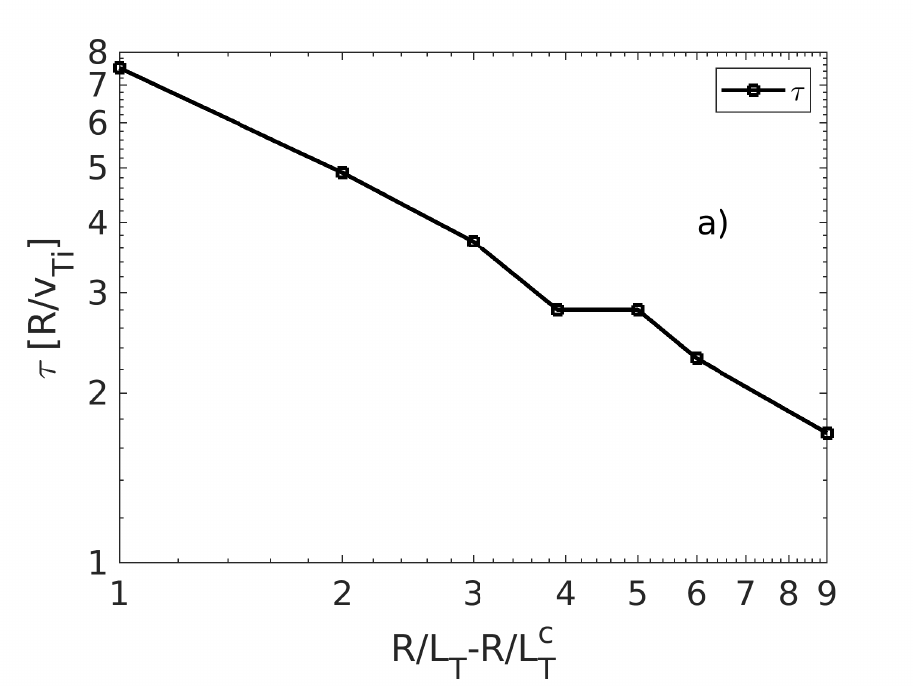}\includegraphics[width=0.5\linewidth]{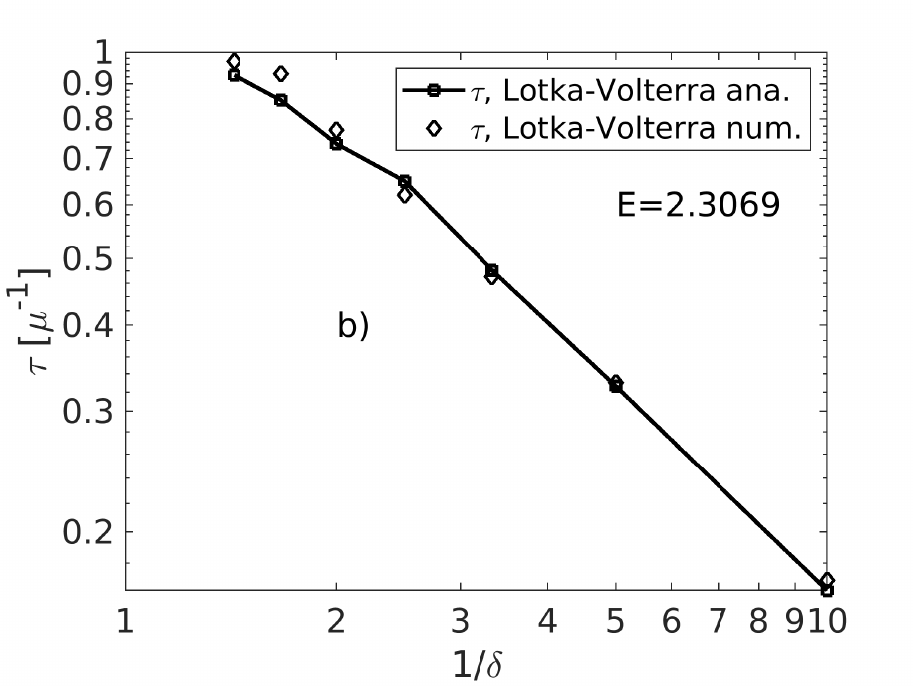}
\caption{Comparison of response times: a) response-time between particle flux $\Gamma_e$ and zonal density energy $n_{ez}^2$ v.s. distance to threshold, and b) response-time between predator and prey in the Lotka-Volterra model.}
\label{fig-resptime}
\end{figure}

The response time is usually defined as the time-lag between maxima \emph{or} minima of two signals. Here, for clarity, we define it as the time-lag between the first maximum of the two signals, as indicated by black arrows in Fig \ref{fig-ctem-sim}a . The response-time $\tau$ is shown v.s. the distance to threshold $R/L_T-R/L_T^c$ \cite{LeiQi-inprep-2021}, related to the linear growth-rate $\gamma_{\rm TEM}$ via $\gamma_{\rm TEM} \sim \sqrt{ \Big| \frac{R}{L_T}-\frac{R}{L_T^c} \Big| }$ [Fig. \ref{fig-resptime}a]. Here, the critical gradient is $R/L_T^c=3$.

In the Lotka-Volterra model, the response time is given analytically by:
\begin{eqnarray}
\tau(\delta, E) = G_{LV}^{-1}(Y_{max}) - G_{LV}^{-1}(\frac{1}{\delta}),
\end{eqnarray}
where $G_{LV}^{-1}(Y,E)$ is the second Lotka-Volterra integral (branch $j=-1$) given by expression (\ref{def-LV-integral1}).

For the Lotka-Volterra PP model, the response time is shown v.s. the parameter $1/\delta$ (i.e. $\gamma / \mu$), at fixed energy $E=3-\ln 2$ [Fig.\ref{fig-resptime}b]. The analytical  result is shown (full symbols) and is compared to the numerical result (open symbols) obtained by solving numerically the Lotka-Volterra Eqs. (\ref{ppx1},\ref{ppy1}). The agreement is reasonable.
One observes that both response times Fig. \ref{fig-resptime}a and \ref{fig-resptime}b increase with increasing drive. Conversely, both response time decrease with decreasing drive. This is consistent with a critical exponent behavior: the response time increases as marginal  stability is approached, i.e. as $R/L_T \to R/L_T^c$ ($L_T=L_{Te}$) in the gyrokinetic simulation and $1/ \delta \to 0$ in the Lotka-Volterra model. Writing $\tau = |R/L_T - R/L_T^c|^{- \alpha}$ and $\tau_{LV} = (\frac{1}{\delta})^{- \beta}$, the following scalings or obtained: $\alpha=0.67$ and $\beta =0.9 $. This trend is also predicted in the traffic-jam model for avalanches of Ref. \cite{Kosuga2014} for a 1D model of zonal ion temperature corrugations. Based on this model, the scaling of a characteristic response time was given as: $ \frac{1}{2 \tau} \sqrt{ \frac{c_0^2 \tau}{\chi_2} } \simeq \gamma_{\rm max}$, where $\gamma_{\rm max} \sim V_{E\times B}'$, with $V_{E\times B}'$ the zonal flow shearing rate, $c_0$ the initial avalanche speed and $\chi_2$ the heat diffusivity. Assuming that $\chi_2$ scales like $\chi_2 \sim |R/L_T - R/ L_T^c|^\eta$, as for stiff transport with $\eta>0$ a scaling exponent, this predicts that the response-time of Ref. \cite{Kosuga2014} scales like $\tau \sim |R/L_T - R/ L_T^c|^{-\eta}$. 

\section{Discussion and conclusions}
Let us first discuss the analytical solution of the Lotka-Volterra model. To our knowledge, this is the first time that such a closed-form exact solution of the model was obtained. 
This analytical solution (\ref{def-LV-integral1},\ref{anasol-predator},\ref{anasol-prey},\ref{def-LV-integral2}) may possibly be used to fit experimental data of limit cycle oscillations in fusion devices. It provides a simpler alternative compared to the method used in Ref. \cite{SKobayashiGurcanDiamond2015}, which extracted directly predator-prey coefficients from gyrokinetic simulation data by solving the Lotka-Volterra system numerically. More precisely, our method extracts two fitting-parameters: the ratio of linear predator damping-rate to prey growth-rate $\delta = \mu / \gamma$, and the `energy' $E$ associated to the limit-cycle. This can be used by the experimental community to know if a signal is `predator-prey', and to extract its key parameters. We sketch briefly how one could proceed: From the experimental prey \& predator signals $X(t)$ and $Y(t)$, one could minimize the squared-difference between $X(t)$ and $\lambdajprey(t)$, and between $Y(t)$ and $\lambdajpreda(t)$ - i.e. a least-square fit - with respect to the fitting-parameters $\delta$ and $E$.
\\
There are some limitations to our model. First, this model differs from the drift-wave zonal-flow model of Ref. \cite{Diamond1994} in that the self-damping term is neglected in the turbulence evolution equation. Although this term is important to obtain steady saturated states in the long time limit, and associated bifurcation between states, this parameter has no effect on the transient limit-cycle regime - apart from displacing its center - on which we focus here. 
Second, we acknowledge that we did not formally invert Eqs. (\ref{def-LV-integral1}) and (\ref{def-LV-integral2}) by solving e.g. Eq. (\ref{predator2}) for $Y$. Instead, we inverted them graphically by plotting $X$ v.s. $F_{LV}$ and $Y$ v.s. $G_{LV}$. The inversion could be easily accomplished by writing the Lotka-Volterra integrals (\ref{def-LV-integral1},\ref{def-LV-integral2}) as a Taylor series, and then  inverting this series, using the Taylor series inverse-function formula. This is left for future work.
Third, we verified our analytical solution only for one value of the energy $E$ and associated initial conditions $(X_0,Y_0)=(2,1)$. It would be useful to verify it for different initial conditions (different energies), but this is beyond the scope of this article.
Let us now discuss the comparison of the response-time between the Lotka-Volterra model and gyrokinetic CTEM simulations.
In Ref. \cite{EJKimHollerbach2020}, a stochastic version of the Kim-Diamond model was studied. The temporal cross-correlation between zonal flow energy and turbulence energy was computed and the associated response-time was obtained (Fig 1c in this reference). Ref. \cite{Kosuga2014} predicted -in the framework of the `traffic-jam' model- that the response-time scales like $\tau \sim |R/L_T - R/ L_T^c|^{- \eta}$, i.e. the response time increases as marginal stability is approached. We showed that a similar trend, i.e. $\tau$ increasing when the prey drive $~1/ \delta$ decreases is obtained when applying the Lotka-Volterra predator-prey model to the interaction between zonal density corrugations and turbulence. This is also consistent with Ref. \cite{DifPradalier2017}, where the probability of finding large scale (and thus slowly evolving) zonal structures is maximal near-marginality (cf. Fig 4 in the latter Reference). More connection between the zonal staircase generation and predator-prey modeling would be interesting for future work.\\
In conclusion, we derived analytically the solutions to the well-known Lotka-Volterra Predator-Prey model. We applied the newly-found solutions to calculate analytically the Predator-Prey response time, and we compared its scaling with that of gyrokinetic simulations of CTEM turbulence. As the Lotka-Volterra model is one of the simplest model to describe self-organized systems, we believe that having an analytical insight into the dynamics of this model is the first step to allow a better understanding of more complicated models.

\section*{Acknowledgements}
The authors would like to thank P.H. Diamond, Y. Kosuga, Raghvendra Singh, X. Garbet, T. Kobayashi, Min-Jun Choi, Jae-Min Kwon and Sumin Yi for helpful discussions. We also thank the anonymous Referees for valuable comments.
This work was supported by R \& D Program through Korean Institute of Fusion Energy (KFE) funded by the Ministry of Science and ICT of the Republic of Korea (KFE-EN2141-7). The data that support the findings of this study are available from the corresponding author upon reasonable request.


\section*{Appendix A: Analogy with Jacobi elliptic functions}

First note that harmonic functions $\sin(\theta)$ and $\cos(\theta)$ can be defined as the inverse of integrals, via:
\begin{equation} \tag{A1}
\theta = \int_0^x \frac{dx'}{\sqrt{1-x'^2}}, \quad {\rm and} \quad \theta = \int_x^1 \frac{dx'}{\sqrt{1-x'^2}},
\end{equation}
respectively. Jacobi elliptic functions arise from a generalization of these formulas, when the trigonometric circle is generalized to an ellipse (although historically, Jacobi discovered these functions using a different approach).

One can make an analogy between the solutions $\lambdajpreda(t,E)$, $\lambdajprey(t,E)$, with $j=0,-1$ and the Jacobi elliptic functions sn(t,k) and cn(t,k), respectively.
Note that Jacobi elliptic functions appear naturally in describing the non-circular bounce/transit periodic motion of guiding centers in tokamak geometry (e.g. the `banana' orbits, etc \ldots) and its applications to residual zonal flow problem \cite{Duthoit2014}.
One way to define the Jacobi elliptic functions is to view them as extended trigonometric functions \cite{Schwalm2015}. Consider an ellipse for which:
\begin{equation} \tag{A2}
\frac{x^2}{a^2} + \frac{y^2}{b^2} = 1
\end{equation}
Defining normalized variables $X=x/b$ and $Y=y/b$, one obtains, after some algebra:
\begin{equation} \tag{A3}
\frac{1}{\kappa} X^2 + \kappa Y^2 = \kappa
\label{ene-ellipse}
\end{equation}
Here $\kappa = a/b = E/b^2$ is the elongation, where $E=ab$ is the energy, proportional to the surface area of the ellipse, i.e. the action invariant, $I= \oint p dq = \pi a b$, where $q=X$ and p=Y are the usual position and momentum conjugate variables. Hence elongation can also be viewed as `normalized energy'.
One clearly sees the analogy between Eq. (\ref{ene-ellipse}) - which represents limit cycles of normalized energy  $E/b^2=\kappa$ - and Eq. (\ref{hamilton1}) for the Lotka-Volterra system. Both represent energy conservation. The only difference is the topology of the limit-cycle, and hence the form of the energy integral. Note also that in both cases, the shape of the  limit-cycle depends on the value of the energy.

Dividing by $\kappa$ and defining the \emph{elliptic modulus} $k = \sqrt{1- \frac{1}{\kappa^2}}$, this can be written:
\begin{equation} \tag{A4}
(1-k^2) X^2 + Y^2= 1
\label{ene-ellipse2}
\end{equation}
 It is well-known that Eq. (\ref{ene-ellipse2}) describes phase-space contours associated to the solutions $X/ \kappa = {\rm cn}(t,k)$ and $Y={\rm sn}(t,k)$, where ${\rm cn}(t,k)$ and ${\rm sn}(t,k)$ are Jacobi elliptic functions, and $\kappa=1/\sqrt{1-k^2}$.

\section*{Appendix B: Expression for the integration constants $T_{Y_{\rm min}}$ and $T_{X_{\rm min}}$}
The quantity $T_{Y_{\rm min}}$ takes the form \cite{PMasson2021}:
\begin{equation} \tag{B1}
T_{Y_{\rm min}}
=
\begin{cases}
- \int_{Y_0}^{Y_{\rm min}} \frac{dY}{Y \Big[ 1+ W_0(-Y^{-1/ \delta} e^{Y-E}) \Big]} , \quad X_0 \le 1, \\
T_{Y_{\rm lower}} - \int_{Y_0}^{Y_{\rm max}} \frac{dY}{Y \Big[ 1 + W_{-1} (-Y^{-1/ \delta} e^{Y-E}) \Big]} , \quad X_0 > 1,
 \end{cases}
 \label{ap-tymin}
\end{equation} 
where $T_{Y_{\rm lower}} = \int_{Y_{\rm min}}^{Y_{\rm max}} \frac{dY}{Y \Big[ 1
+ W_{-1} (-Y^{-1/ \delta}) e^{Y-E} \Big]}$ is the lower semi-period for the predator population.
Similarly, the quantity $T_{X_{\rm min}}$ reads:
\begin{equation} \tag{B2}
T_{X_{\rm min}}
=
\begin{cases}
T_{X_{\rm upper}} + \int_{X_0}^{X_{\rm max}} \frac{dX}{ X/ \delta \Big[ 1 + W_0 (-\delta X^{-\delta} e^{\delta (X-E)}) \Big]} , \quad Y_0 \le \frac{1}{\delta}, \\
\int_{X_0}^{X_{\rm min}} \frac{dX}{X / \delta \Big[ 1+ W_{-1}(- \delta X^{- \delta} e^{\delta (X-E)}) \Big]} , \quad X_0 \le 1,
 \end{cases}
 \label{ap-tymin}
 \end{equation}
 where $T_{X_{\rm upper}} = - \int_{X_{\rm min}}^{X_{\rm max}} \frac{dX}{X / \delta \Big[ 1
+ W_{-1} (- \delta X^{- \delta}) e^{\delta (X-E)} \Big]}$ is the upper semi-period for the prey population.
Note that similar formulas are given in Ref. \cite{PMasson2021} for the quantities $T_{Y_{\rm upper}}$ and $T_{X_{\rm lower}}$. The exact period $T_{\rm LCO}$ of the orbit (or limit cycle) is then given by the sum  of the two semi-periods:
\begin{equation} \tag{B3}
T_{\rm LCO} = T_{X_{\rm upper}} + T_{X_{\rm lower}} = T_{Y_{\rm lower}} + T_{Y_{\rm upper}}
\end{equation}

\end{document}